# Insider Threat Detection Based on Stress Recognition Using Keystroke Dynamics

Azamat Sultanov, Konstantin Kogos

Institute of Cyber Intelligence Systems, Cryptology and Cybersecurity Department

NRNU Mephi, Moscow 115409, Russian Federation

sae010@campus.mephi.ru, kgkogos@mephi.ru

## ABSTRACT

Insider threat is one of the most pressing threats in the field of information security as it leads to huge financial losses by the companies. Most of the proposed methods for detecting this threat require expensive and invasive equipment, which makes them difficult to use in practice. In this paper, we present non-invasive method for detecting insider threat based on stress recognition using user's keystroke dynamics assuming that intruder experiences stress during making illegal actions, which affects the behavioral characteristics. Proposed method uses both supervised and unsupervised machine learning algorithms. As the results show, stress can provide highly valuable information for insider threat detection.

## INTRODUCTION

In the modern world of information technology, ensuring information security is becoming an increasingly difficult task. Despite the enormous efforts of companies developing means of effectively countering information threats, for some of the threats this task remains unresolved.

One of the threats that most companies face is the insider threat. This type of threat poses the greatest danger due to the huge number of sources generating it and the lack of effective means to counter this threat, as a result of which companies suffer huge financial losses. That is why the insider threat or an internal intruder threat is relevant and at the moment it is receiving considerable attention from researchers in the world of information security.

The explanation for the lack of sufficient means to counter the threat of an internal intruder is the nature of this threat. Even though the result of the actions of the internal intruder is obvious, it is very difficult to find informative indicators that would allow to detect abnormal behavior of the user of the information system and distinguish it from normal. To analyze the behavior, one can use the data of the electrocardiogram (ECG) or electroencephalogram (EEG) [1,2,3], body temperature [1], skin conductivity [2], eye movement [4] and other biometric indicators.

However, the methods for obtaining most types of biometric indicators are invasive, that is, they require the use of special and expensive equipment in the form of sensors and cameras, which are most often in direct contact with the experiment participant in the process of data collection. This drawback is the reason that methods based on the usage of invasive equipment for obtaining biometric indicators are difficult to put into practice.

At the same time, there are some non-invasive methods of accumulating data for behavioral analysis, which are based on the use of very affordable tools, such as keyboards and mice, which can be easily found in any office. These low-cost tools can act as sensors that provide behavioral information or, in other words, a behavioral characteristic consisting of keyboard handwriting, keystrokes, and gestures. Behavioral characteristics can be used in solving authentication problems [5,6,7,8,9], emotional state detection [10,11,12,13,14,15], as well as in tasks of detecting the threat of an internal intruder [16].

This work is devoted to the study of methods for detecting an internal intruder by identifying the stressful state of the user based on the analysis of the interaction with the keyboard and mouse. As in [16], the study is based on testing the assumption that behavioral indicators change when an unlawful act is committed by an internal intruder under the influence of an induced stress state. However, unlike the work [16], a wider range of features extracted from raw data is used and a larger number of algorithms are used to detect threats.

The first section analyzes the related works. First, a class of works based on the use of supervised machine learning algorithms is analyzed, the data sets used are considered within each work, and the classification results of the applied algorithms are presented. Similarly, a class of works using unsupervised machine learning algorithms is considered. Methods for detecting the threat of an internal intruder by identifying stress are also considered.

The second section presents the process of data collection, which includes the requirements for scenarios for data collection, proposed scenarios, their description and justification for the use of these scenarios.

The third section describes the process of extracting features from the log files, their processing and preprocessing.

In the fourth section, models of classifiers of abnormal and normal behavior, based on supervised machine learning algorithms are considered, the results of models evaluations are presented, and their comparative characterization is carried out.

In the fifth section, anomaly detection models based on unsupervised machine learning algorithms are considered, the results of evaluations of these models are presented, and their comparative characteristics are carried out.

## 1  Related Works

As the [17] states, the threat of an internal intruder may come from an active or former employee or business partner who has or had privileges to access the network, system or data of the organization and who intentionally or unintentionally perform actions that adversely affect the confidentiality, integrity and accessibility of the organization or information system.

The main types of internal intruders, as provided in [18], are those who act intentionally and maliciously and those who pose a threat to the organization due to their inattention, negligence and the commission of inadvertent actions. As it is also stated in [18], the main goals and objectives of the first category include sabotage, intellectual property theft, espionage and financial fraud.

Causes of threats from members of the second category are human errors, phishing, malware, unintentional aiding and stolen accounts.

The threat of an internal intruder is considered as the main security problem of organizations [19] and according to the data given in [20] approximately 87% of cases of threats to the security of an organization are recorded as threats from internal intruders. In this paper, we will consider only cases of intentional creation of a security threat by an internal intruder and provide the main detection methods using machine learning algorithms.

Three main categories of machine learning algorithms are used to detect the threat of an internal intruder:

- supervised learning algorithms;
- unsupervised learning algorithms;
- anomaly detection algorithms from the group of unsupervised learning algorithms;

## 1.1 Insider Threat Detection Using Supervised Learning Algorithm

In [21], the authors conduct a comparative analysis of classifiers trained on the publicly available CERT Insider Threat database dataset, which is a collection of a large number of real cases of internal intruder threats with information about web and email resource traffic, as well as file system logs. As the algorithms that form the basis of the classifiers, such as Random Forest, k-NN, Gradient Boosting, Decision Trees, Logistic Regression, their variations and combinations were chosen. The best results were shown by classifiers based on the Random Forest algorithm and its variations. The achieved accuracy both on a full and on a trimmed dataset ranged from 94% to 98%. However, Random Forest-based classifiers have lost in the speed of training to k-NN-based classifiers, which at best provide prediction accuracy slightly worse than the maximum accuracy in the case of Random Forest.

A similar dataset is analyzed in [22], however, in this work, in addition to the Random Forest, Support Vector Machines, Decision Trees, and Logistic Regression algorithms, neural networks are used. A version using boosting algorithms was also created for each of the models, which made it possible to increase the accuracy of predictions for almost all the models. The results of models based on neural networks did not exceed the accuracy of models based on Random Forest, Decision Trees and Logistic Regression, while the minimum classification accuracy among all classifiers equaled to almost 92%.

The work [23] stands out among the others in that the training of classifier models does not take place on a static dataset, but on a continuous data stream. As an algorithm for the classifier model, One-Class SVM is used, which, after bringing the training sample into the attribute space with a larger dimension, is considered as a regular Support Vector Machine. The method obtained 71% classification accuracy.

The article [24] attempts to create a model of a multiclass classifier based on the k-NN algorithm for solving the two-factor authentication problem.

The classifier predicted that a user would belong to one of four groups based on face recognition, which are legitimate, possibly legitimate, possibly non-legitimate and non-legitimate.

## 1.2 Insider Threat Detection Using Unsupervised Learning Algorithms

This section describes the works that use unsupervised learning algorithms at different stages of data analysis, particularly clustering methods.

In [25], the detection of the threat of an internal intruder is implemented using preliminary clustering of data based on graphs and subsequent identification of anomalies. The CERT database was used as a data source. The model evaluation gave the best AUC score of 0.76.

The authors of [26], use a model based on one of the varieties of deep neural networks, Deep Belief Networks to detect an insider threat. The dataset used is part of the public CERT database. The Deep Belief Network was optimized using the Golden Section Search algorithm, and the result of the network was used to train the classifier based on Support Vector Machines. The proposed method achieved 97.8% accuracy of detecting the threat of an internal intruder.

Several variations of the k-means method are used by authors in [27]. The dataset is taken from the CERT database. The best model for anomaly detection achieved 76.71% accuracy with the share of a 20% rate of false-positive predictions.

### 1.3 Anomaly-based Insider Threat Detection

The most common methods for detecting the threat of an internal intruder are based on anomaly detection algorithms. In these approaches, models are most often trained on data that consists almost entirely or completely of examples with normal behavior, traffic, or other characteristics. Having trained on such data, models are capable of detecting examples that differ greatly in properties, which makes it possible to efficiently identify anomalies, particularly the threat of an internal intruder.

The authors of [28] use neural networks training on vectors transformed using the word2vec algorithm using the UNSW-NB15 dataset that contains real examples of normal and synthetic examples of abnormal traffic. The optimal configuration resulted in a precision of 99.20%, a recall of 82.07%, and a false positive rate of 0.61%.

In the article [29], a dataset from the CERT database is used for research. Standard anomaly detection algorithms have been applied, including Isolation Forest and One-Class Support Vector Machines. Models are constructed in such a way that they analyze the dataset previously divided into parts by time cycles. In the worst case, the AUC metric yielded a value of 0.87, while in all experiments, models based on One-Class Support Vector Machines performed better.

One of the varieties of the CERT dataset (CERT Insider Threat v6.2) is analyzed to identify the threat of an internal intruder using the anomaly detection approach in [30]. Models are based on deep and recurrent neural networks. The best model on average, with a confidence of 95.53%, detects anomalies present in the dataset.

Another example of the use of anomaly detection algorithms to detect the threat of an internal intruder is given in [31]. Authors use auto-encoders, analyzing the NSL-KDD dataset with information about network connections. The proposed method produced a detection accuracy of 91.70%.

### 1.4 Insider Threat Detection Based on Stress Recognition

There are works in which the insider threat detection is based on the assumption that when committing abnormal actions, a user who is a current or former employee of the organization experiences stress. Stress, in turn, affects biometric indicators such as heart rate, blood pressure, body temperature, as well as the rhythm and dynamics of keystrokes on devices that interact with the information system. Therefore, by identifying stress using biometric indicators, one can build the chain in the reverse order and with a certain degree of confidence assume that the stress was caused due to unlawful actions that constitute an insider threat. Such an assumption was made in [1,2,3,4].

In the paper [3], the authors carry out a comprehensive study, including the processes of data collection, pre-processing, training of models of classifiers and their evaluation. As the main biometric indicators for stress monitoring, electrocardiogram (ECG) or electroencephalogram (EEG) signals were chosen. Of greatest interest is the data collection process, which uses scenarios that describe both the actions of an internal intruder causing a stressful state and the actions of an ordinary employee of a company that do not pose a security risk and do not induce stress. The Support Vector Machines algorithm is selected as the basis of the classifier model. The proposed model achieved 86% prediction accuracy.

A big drawback from the point of view of practical application in [3] is the need for special equipment that tracks changes in biometric indicators. To solve this problem in the article [16], the authors use the keyboard and mouse as tracking sensors, which provide valuable information about the dynamics of keystrokes and buttons. As already shown in [10,11,12,13,14,15], the keyboard script changes under the influence of various emotional states. In [16], four classifier models were constructed based on Support Vector Machines, Random Forest, k-NN, and Bootstrap Aggregating. Models were trained on data collected using two categories of scenarios, as in [3], describing the actions of an internal intruder and an ordinary employee. The resulting model accuracy ranged from 67.5% to 72.5%. The authors attribute the poor results to a lack of informative features isolated from the dynamics of using the keyboard. Most of the signs were isolated from the dynamics of using the mouse, but the conclusions noted that the influence of stress on the dynamics is obvious, and this confirms the assumption that the ability to detect stress through biometric indicators allows you to detect the insider threat.

The current study is in many respects similar to [16], however, there are significant differences in the processes of data accumulation and feature extraction, as well as in the used machine learning algorithms for classifier models and the additional use of anomaly detection algorithms. Further in the work, the terms stress and anomaly will be considered interchangeable to follow algorithmic terminology.

## 2   Data Collection Process

Due to the lack of datasets for the current study, it became necessary to organize data collection and create a dataset for training of classifier models and anomaly detection models. To do this, two categories of scenarios were invented that describe the actions of an internal intruder and the actions of a non-threatening employee.

Scenarios of the first category were designed in such a way that stress was induced in the experiment's participant during their execution. For this, the scenarios of the first category, unlike the scenarios of the second category, were limited in time, and the actions performed by the participants of the experiment as an internal intruder described real situations of attacks that could arise in organizations.

In the first scenario of this category, an internal intruder (employee or former employee) sits at the computer of another employee. Finds an archive with confidential salary information in the company. The archive is password-protected, but the attacker is well acquainted with the owner of the computer and has a list of possible passwords, one of which is guaranteed to be suitable. The unpacked archive contains information on the salary of each employee, which is saved as a screenshot. The intruder collects as much data as possible from the screenshots, combines them into a text document and sends them to the provided list of users by mail, entering a description and a subject. Wherein, it is forbidden to send screenshots instead of a text file, since the company's network has a service that monitors all outgoing files over 20 kb. If possible, the intruder clears the browser history.

In the second scenario of this category, the internal intruder (employee) sent a request for the salary increase to the chief in the form of a document. The chief read the letter but did not print it. While he is not in the workplace, the attacker tries to change the contents of the document on the chief's computer, which he left turned on, adding additional items to the form. The intruder enters the chief's mail, finds his letter, downloads the document, deletes the letter, changes the contents of the document, enters his mail and sends a new letter with the changed document to the chief through another browser. Marks the letter in the chief's mailbox as read. If possible, intruder cleans history.

In the third scenario, the intruder is the chief designer at the processor manufacturing company. On his computer, drawings and characteristics of the new processor are stored. An attacker needs to send data to competitors, but it is strictly forbidden to do this by copying data, creating a new document, and sending by mail, as the company has a system to prevent information leaks. However, the intruder is aware of the flaws of the system. Every 3 hours for 4 minutes, the system archives the collected data, and the keylogger turns off. Using this resource to temporarily store text, an attacker needs to retype the characteristics there and publish. several posts on a page in a social network, wherein each post there will be a part of the link to the characteristics stored on the resource.

The scenarios of the second category were not limited in time. These scenarios described the actions that an employee in the organization could perform while doing his job.

In the first scenario, the employee needs to find answers to the questions provided in the list in the browser. Reprint the answers into the created document. Next, enter the mailbox using the credentials provided during the experiment. The email address and password are selected so that they are easy to enter to simulate an employee who often uses a mailbox and knows his email address and password by heart. It is necessary to create a new letter by attaching a file with the answers, enter the subject and description of the letter, a list of recipients with electronic addresses of varying complexity and send it.

In the second scenario of this category, an employee works in a company with a large number of branches. The reporting of branches is compiled at the head office, and the employee, considered in the scenario, is engaged in this process. Last month's reports came to him in the form of tables. It is necessary to download all the documents from the mailbox, compose them so that the generalizing document must contain the fields: branch, date, income and expenses. In the last line, the employee calculates the sum of the fields of income and expenses. Wherein, it is forbidden to copy data from branch reports and all data must be entered manually. After the completion of the general report, the employee sends it by mail to the same branches with an arbitrary subject and body of the letter.

In the third scenario, an event management company's employee received a letter asking him to send information about the company's achievements over the previous 2 years to participate in a tender for an annual hackathon for biologists. Information is stored on the employee's computer, but is scattered in several files. It is necessary to collect the required information from the local storage and send it in a response letter, after combining the collected data in the form of a document. The information must be reprinted, not copied.

Due to some circumstances, only 8 people took part in the data collection experiment. All participants were students with an average age of 20 years. Wherein, all the participants executed scenarios on the same device in order to avoid the influence of keyboard and mouse types on data.

Python software was written to log keyboard and mouse events during the execution of the scenarios.

### 3   Feature Selection and Preprocessing

To analyze the data on the interaction of the user with the keyboard and mouse, most often time and frequency features are distinguished, as in [10,11,14,16]. The time features that are used in the current study are graphically presented in Figure 1.

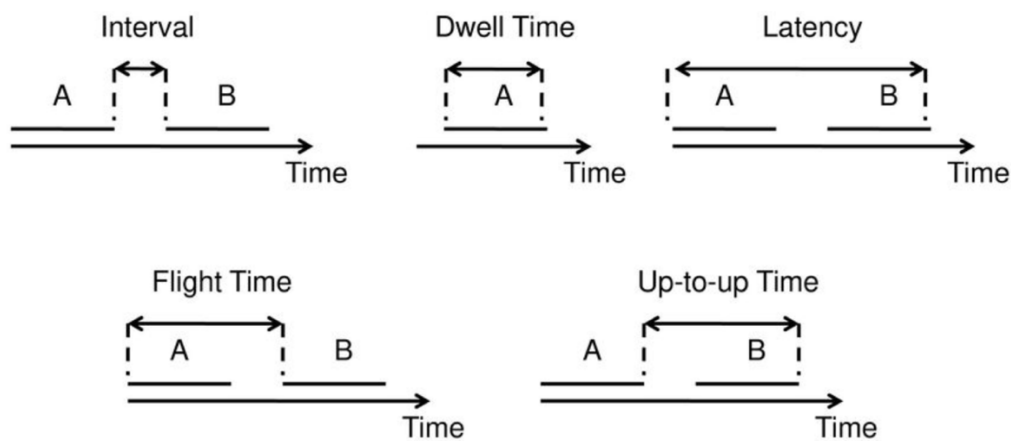

Figure 1 — schematic representation of time features [32]

Lists of time and frequency features are presented below in Tables 1 and 2.

Table 1 — time features

| № | Name | Definition |
|---|------|------------|
| 1 | dwell time | time (in milliseconds) between pressing and releasing the same key |
| 2 | flight time or down-to-down | time (in milliseconds) between pressing one key and pressing another key |
| 3 | latency | time (in milliseconds) between pressing one key and releasing another key |
| 4 | interval | time (in milliseconds) between releasing one key and pressing another key |
| 5 | up-to-up | time (in milliseconds) between releasing one key and releasing another key |

Table 2 — frequency features

| № | Name | Definition |
|---|------|------------|
| 1 | typing speed | keys pressed or typed words per minute |
| 3 | frequency of presses | key usages per minute |

In contrast to the work [16], where the features were extracted more for the mouse, in the current study, time and frequency features, in addition to the typing speed, are extracted for four large groups:

- mouse clicks;

- special keys like backspace, del, capslock, shift, tab, alt;

- bigrams, which are a combination of two letters of the alphabet;

- trigrams, which are a combination of three letters of the alphabet.

The features for the bigrams and trigrams were extracted only for the Cyrillic alphabet since almost all scenarios required the use of the Russian layout most of the time, and the Latin layout was used in extremely rare cases and not in all scenarios, which made the features for the bigrams and trigrams of the Latin alphabet uninformative due to their complete or almost complete absence in the event logs.

Bigrams and trigrams were not chosen randomly. Selected combinations are the most common in Russian. About 275 million characters of text in Russian were used to calculate the frequencies of bigrams and trigrams [33]. Chosen bigrams and trigrams are presented in the Table 3.

Table 3 — bigrams and trigrams

| Group | Elements |
|---|---|
| bigrams | ст, ен, ов, но, ни, на, ра, ко, то, ро |
| trigrams | ени, ост, ого, ств, ско, ста, ани, про, ест, тор |

Both frequency and time features were calculated using a python script. Using written script, for each of the log files frequency features for special keys and time features for all other groups including special keys were extracted. Before saving to the dataset, for each of the attributes of the final feature vector, except the frequency ones, the average value was calculated within each log file.

The dimension of the feature space in the compiled dataset, taking into account some discarded features, equaled 191. Discarding features was the first step in the data preprocessing process, during which those features that occurred only a few times in all log files were excluded. These features are for right mouse button, caps lock, esc and alt.

At the next stage, those features for whose values the standard deviation calculated separately for the categories of abnormal and normal behavior turned out to be less than the specified acceptable value, were removed.

This is explained by the fact that in the big amount of samples of the dataset there were no values for most of the features, and if they were preserved, algorithms for extracting the most informative features would select features with a standard deviation close to zero. However, these features cannot be used to maximize the predictive ability of models due to the low informational content of such features. As a result of this rejection, the dimension of the feature space was reduced to 150.

Since in the used scenarios the sets of occurring elements from the above groups are different, there were voids for some of the features in the dataset. As the second step of data preprocessing, these voids were filled with medians, which calculated separately for the anomalous and normal categories.

The relative distribution of feature medians for samples with normal and abnormal behavior is shown in Figures 2 and 3.

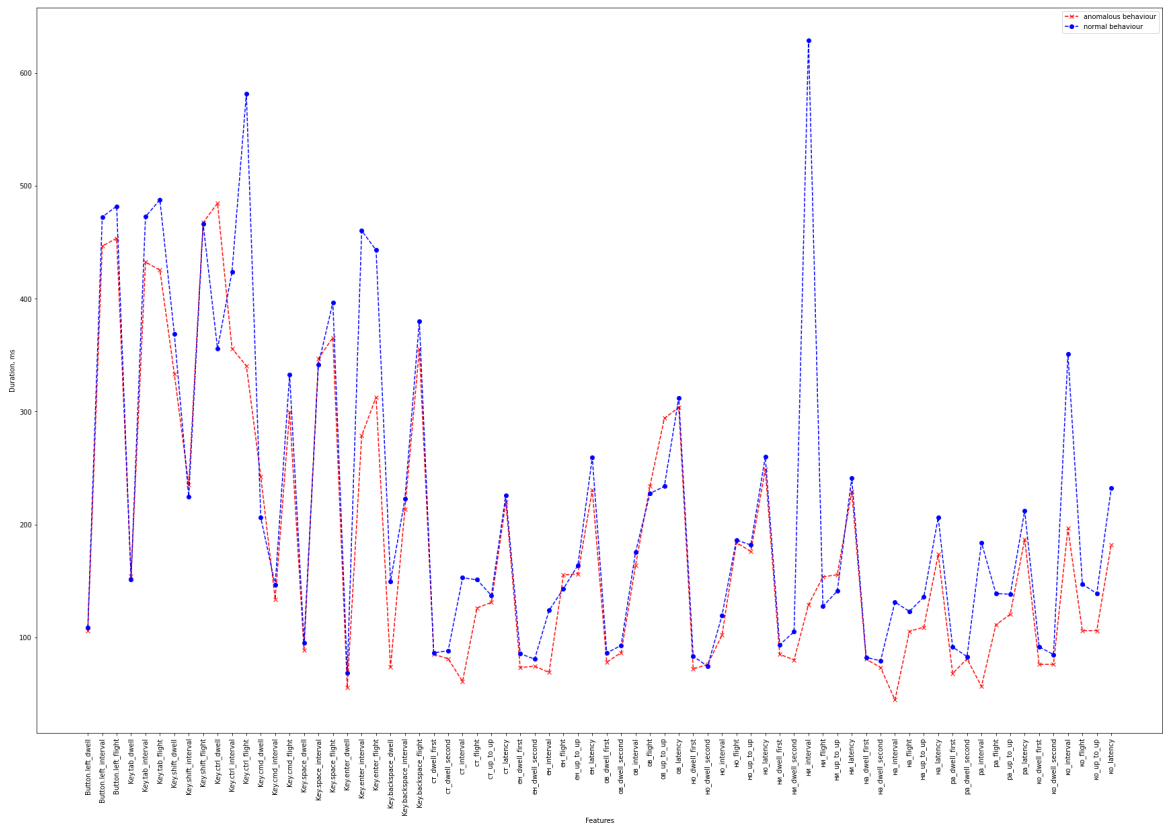

Figure 2 — relative distribution of feature medians, part 1

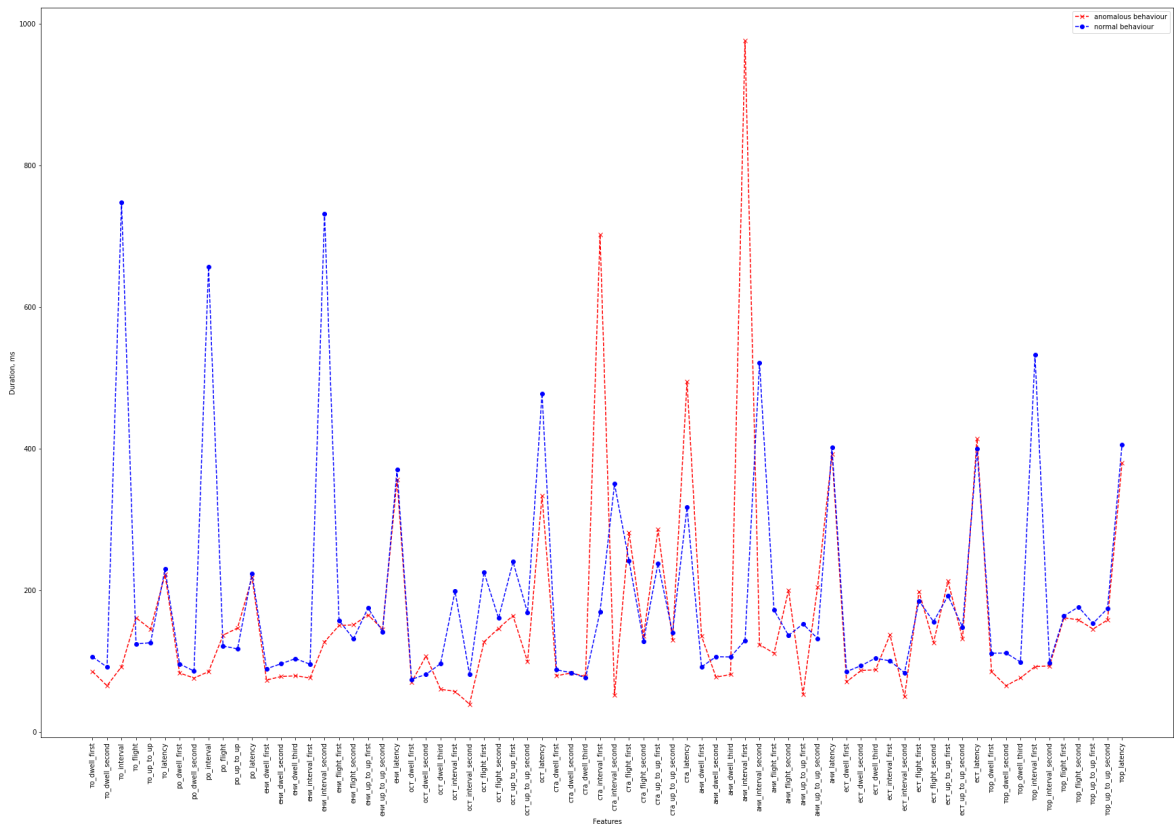

Figure 3 — relative distribution of feature medians, part 2

At the last step of the preprocessing, the Select K-Best algorithm based on chi-squared statistics was applied to extract the three most informative features. This algorithm performs a series of one-dimensional statistical tests, giving each attribute a certain score, according to which sorting and selection of k-best attributes take place. As a result, the following features were selected as the most informative ones:

- то_interval;

- ста_interval_first;

- ани_interval_first.

Reducing the dimension of the space of features to three allowed us to visualize the distribution of dataset samples in the space of features from various angles. This distribution is shown in Figure 4.

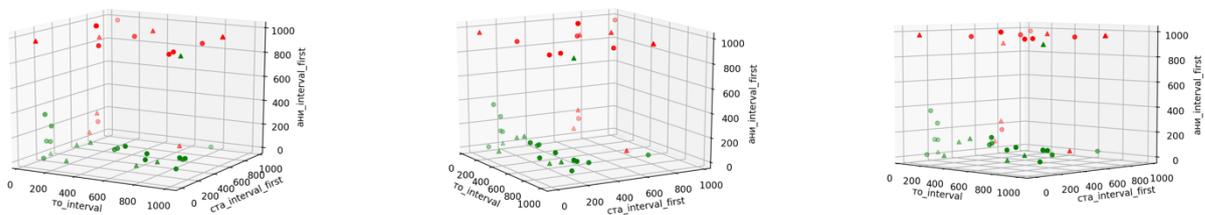

Figure 4 — spatial distribution of samples

## 4  Classifier Models and Their Evaluation

This section summarizes the results of using supervised machine learning algorithms to classify normal and abnormal user behavior.

To construct models of classifiers, the algorithms Logistic Regression, k-NN, Random Forest, Multi-Layer Perceptron and Gradient Boosting were chosen.

The constructed models were trained on 49% of the entire dataset, validated on 21% of the dataset and were tested on the rest of the dataset. The obtained results are presented in Table 4.

Table 4 — evaluation of classifier models

| № | Algorithm | Precision Not-Stress (train / test) | Precision Stress (train / test) | Recall Not-Stress (train / test) | Recall Stress (train / test) | Accuracy (train / test) |
|---|-----------|------------|------------|------------|------------|------------|
| 1 | Logistic Regression | 1.00 / 0.70 | 1.00 / 0.83 | 1.00 / 0.88 | 1.00 / 0.62 | 1.00 / 0.75 |
| 2 | k-NN | 0.92 / 0.73 | 1.00 / 1.00 | 1.00 / 1.00 | 0.91 / 0.62 | 0.95 / 0.81 |
| 3 | Random Forest | 1.00 / 0.80 | 1.00 / 1.00 | 1.00 / 1.00 | 1.00 / 0.75 | 0.91 / 0.88 |
| 4 | Multi-Layer Perceptron | 1.00 / 0.70 | 1.00 / 0.83 | 1.00 / 0.88 | 1.00 / 0.62 | 1.00 / 0.75 |
| 5 | Gradient Boosting | 1.00 / 0.70 | 1.00 / 0.83 | 1.00 / 0.88 | 1.00 / 0.62 | 1.00 / 0.75 |

Table 4 shows that for models based on Logistic Regression, MLP and Gradient Boosting, the same metric values are obtained. Most likely, such a coincidence in the metric values occurred due to the little amount of data and successful spatial distribution of anomalous and normal examples in the space of features.

The latter allowed all models to achieve maximum or near-maximum performance in the training sample. However, models based on Logistic Regression, MLP and Gradient Boosting showed the worst performance in the test sample. In the case of Logistic Regression, this could be explained by the weak predictive ability of this algorithm for the non-linear data distribution, as a result of which the model finds it difficult to catch the pattern of data distribution. And in the case of MLP and Gradient Boosting, such results can be explained only by an insufficient amount of data, since both MLP and Gradient Boosting can capture very complex patterns in the data. Perhaps the performance of the Logistic Regression-based model will be better with increasing data volume, but it is obvious that as the data distribution function becomes more complicated, the proportion of incorrectly predicted samples will increase.

Wherein, the model based on k-NN slightly outperformed the previously mentioned models in terms of performance. Such results can be explained by the fact that k-NN uses the distance between objects in the feature space for classification, and if paid attention to the location of samples with normal and abnormal behavior in Figure 4, it is noticeable how clear is the spatial separation. This separation makes it easy to determine whether an example belongs to a particular class. However, with the increase in the amount of data, new spatial clusters may appear, and the performance of the model based on k-NN can significantly decrease.

The model based on the Random Forest algorithm showed the best results. This is explained by the design of the Random Forest algorithm, which uses a large number of weak classifiers, in this case, decision trees combined into an ensemble, which allows one to consider a large number of functions describing the distribution and averaging the results to obtain a significant increase in comparison with the same decision trees used separately. However, in this case, model overfitting could occur, since on the validation sample the scores are much worse than for models based on Logistic Regression, Gradient Boosting and MLP.

Figure 5 presents the visualized classification results for the train and test samples for the best model. Visualized results for the remaining models are available in Appendix A.

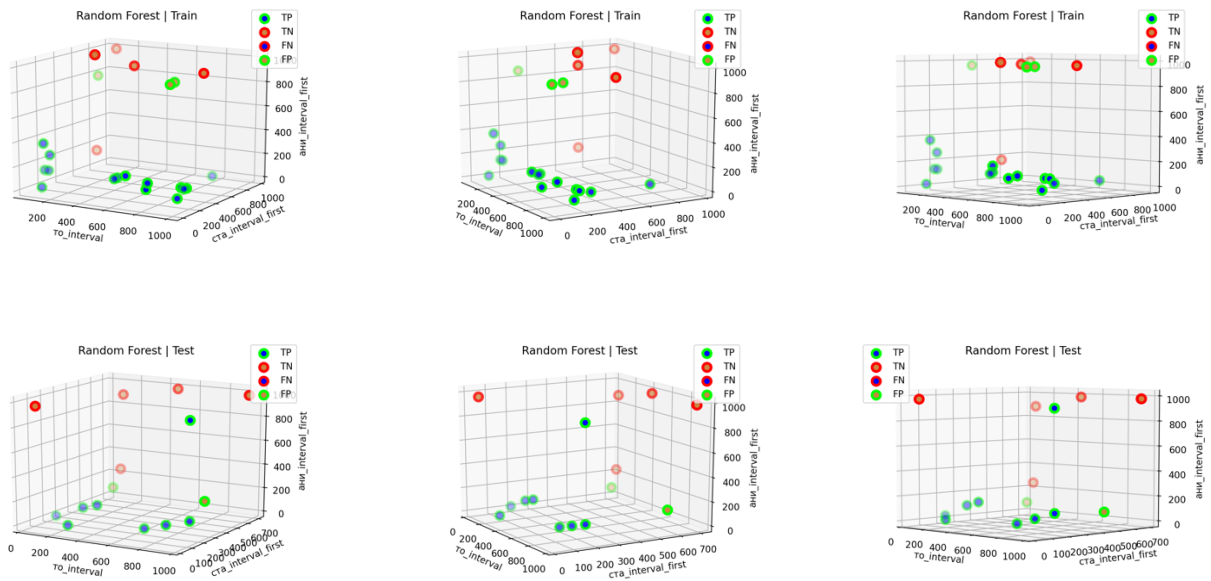

Figure 5 — classification results for the model based on Random Forest

## 5 Anomaly Detection Models and Their Evaluation

This section summarizes the results of applying anomaly detection algorithms.

Algorithms for detecting anomalies are based on two assumptions [29]:

- firstly, most user behavior patterns in the system are normal with a small percentage of abnormal cases;

- secondly, abnormal cases are statistically different from normal.

Besides the fact that there is no need to label data for model training, the main advantage of using anomaly detection algorithms is the ability to adapt to new patterns of abnormal behavior.

To construct anomaly detection models, the algorithms Isolation Forest, One-Class Support Vector Machines, Robust Covariance and Local Outlier Factor were selected. Models are trained only on normal samples, the number of which is 33% of the total dataset. Models are tested on all the abnormal and the remainder of the normal samples. The results of training and testing are shown in Table 5.

Table 5 — evaluation of anomaly detection models

| № | Algorithm | Precision Not-Stress (train / test) | Precision Stress (train / test) | Recall Not-Stress (train / test) | Recall Stress (train / test) | Accuracy (train / test) |
|---|-----------|------------------|------------------|------------------|------------------|------------------|
| 1 | Robust Covariance | 1.00 / 0.60 | — / 0.91 | 0.81 / 0.75 | — / 0.83 | 0.81 / 0.81 |
| 2 | One-Class SVM | 1.00 / 0.00 | — / 0.75 | 0.81 / 0.00 | — / 1.00 | 0.81 / 0.75 |
| 3 | Isolation Forest | 1.00 / 1.00 | — / 1.00 | 0.81 / 1.00 | — / 1.00 | 0.81 / 1.00 |
| 4 | Local Outlier Factor | 1.00 / 0.88 | — / 0.96 | 0.88 / 0.88 | — / 0.96 | 0.88 / 0.94 |

The model based on One-Class SVM showed the worst results without identifying any samples with normal behavior in the test sample. This can only be explained by the fact that the hyperplane separating the class with normal behavior was incorrectly located during the training.

The remaining models showed results much better than the one based on One-Class SVM. The model based on the Local Outlier Factor in the test set showed results close to ideal. This is because the samples are well divided in the feature space, and since the idea of the Local Outlier Factor algorithm is to use spatial characteristics with respect to k-nearest neighbors, in this case, the algorithm was able to easily determine the location of the group of samples with normal behavior.

The model based on the Isolation Forest showed ideal metrics in a test set. And the possible explanation is that Isolation Forest managed to separate the samples as accurately as possible in the attribute space, recursively sorting through all possible variants of spatial sections.

Figure 6 presents the visualized results of the model based on the Isolation Forest algorithm. Visualized results for the remaining models are available in Appendix B.

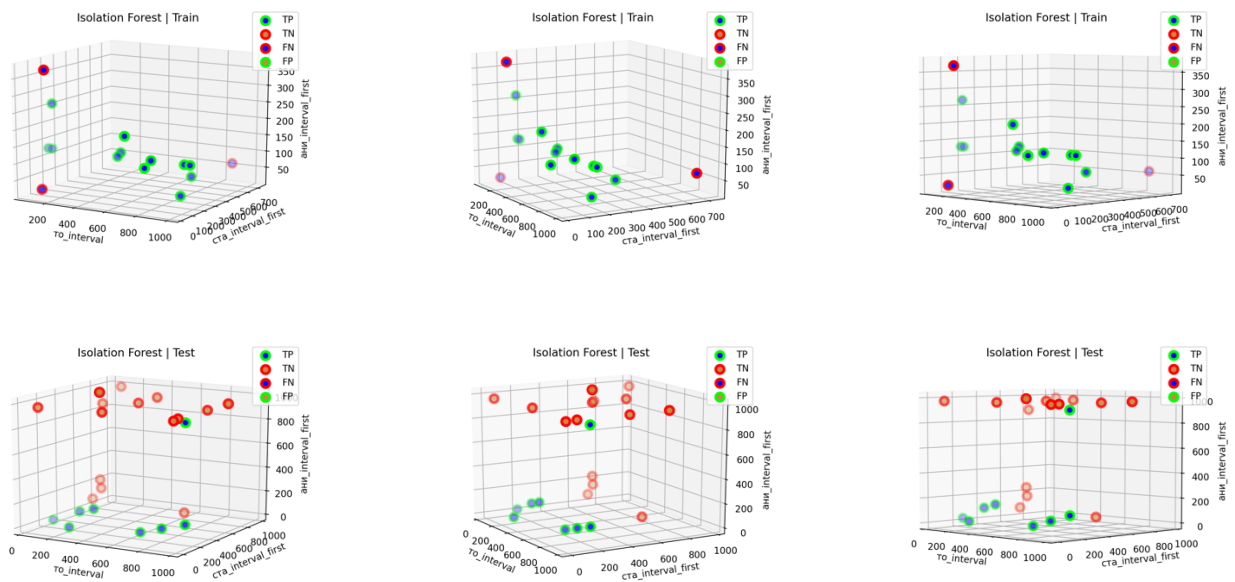

Figure 6 — anomaly detection results for the model based on Isolation Forest

## Conclusions and Further Research

In this paper, an analysis of studies in the field of detecting the insider threat using various biometric indicators was carried out, and the possibility of detecting anomalies by detecting stress using data from interaction with the keyboard and mouse was demonstrated.

In the course of the study, a dataset including the data of the keyboard and mouse logs was collected. The data collection experiment involved 8 people who completed scenarios that described both the actions of a normal employee and the actions of an internal intruder.

Then, using the written software, groups of time and frequency features were extracted from raw data. Steps to preprocess the data and extract the most informative features were also taken. Classifier models based on 5 different supervised machine learning algorithms and anomaly detection models based on 4 unsupervised machine learning algorithms were built. The best classifier model achieved 88% accuracy. The best anomaly detection model achieved 100% accuracy.

The study showed that under the influence of stress, the dynamics of using the keyboard and mouse changes significantly, which leads to an increase in the speed of pressing keys and buttons. The results obtained when testing the models of classifiers, as well as models for detecting anomalies, confirmed the assumption that stress detection can play a key role in detecting anomalies, one of the reasons for which may be an internal intruder.

In the further study, it is planned to collect much more data from different age categories of users with different keyboard usage abilities, which will allow us to analyze all the events on the keyboard and mouse when selecting features that were omitted in this study.

# Appendix A
## Visualized Classifier Models Results

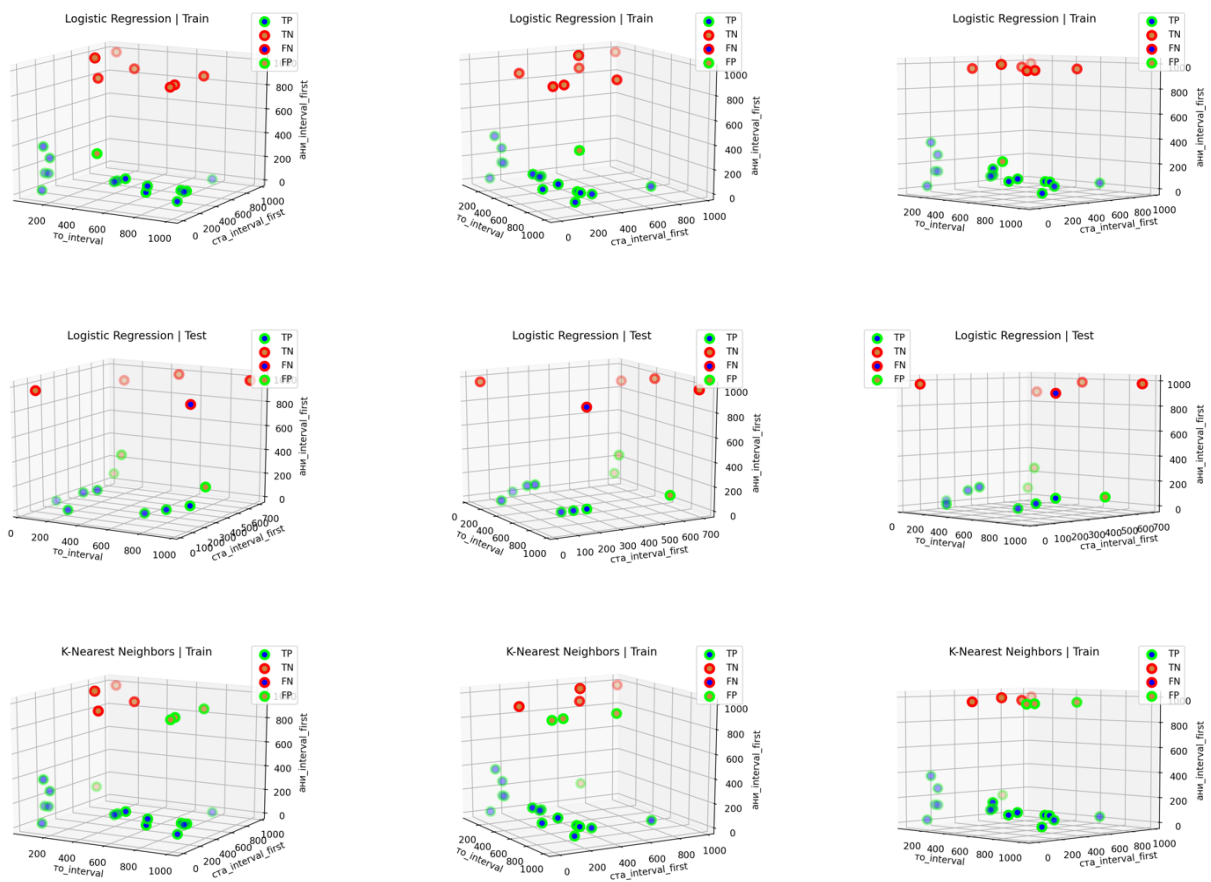

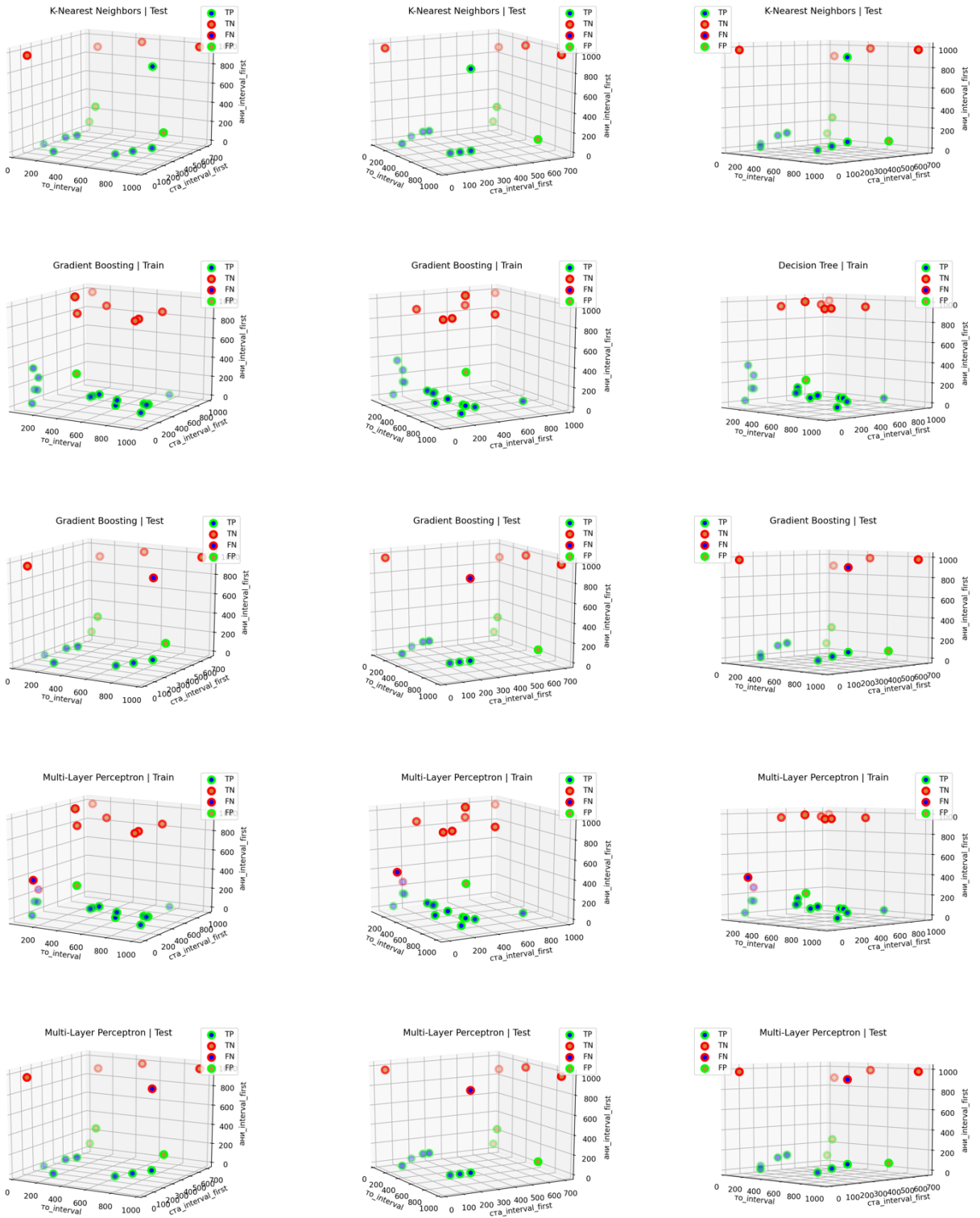

# Appendix B

## Visualized Anomaly Detection Models Results

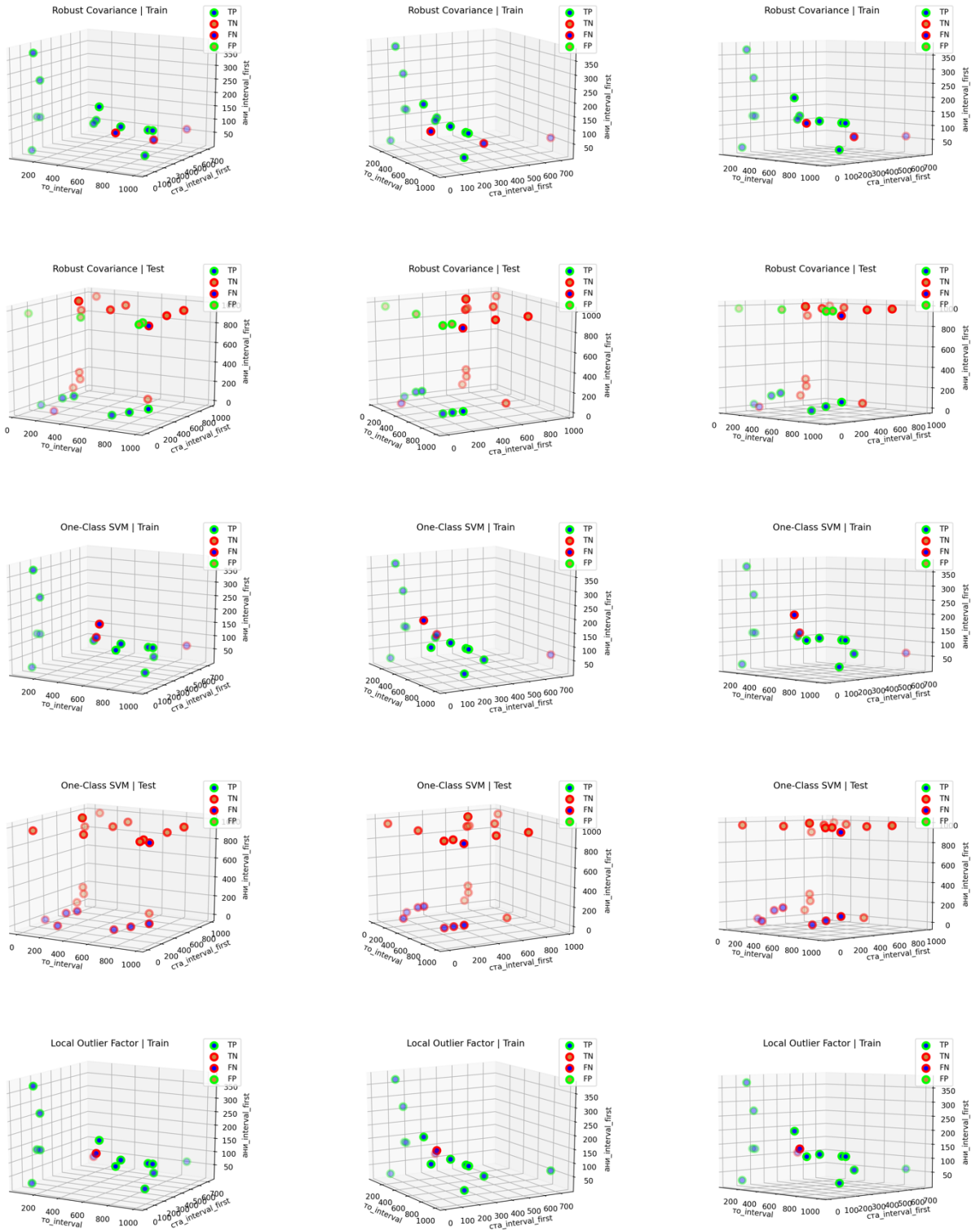

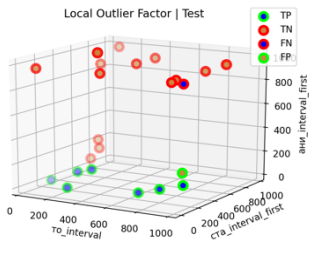 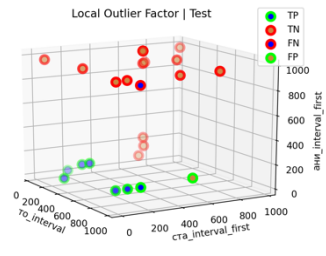 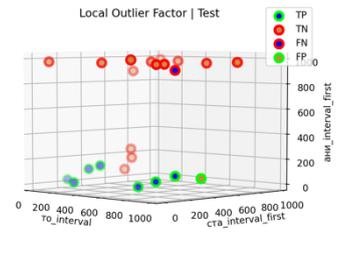